\documentclass[aps,superscriptaddress,showpacs]{revtex4-2}
\usepackage[utf8]{inputenc}
\usepackage{color}
\usepackage{babel}
\usepackage{array}
\usepackage{units}
\usepackage{bm}
\usepackage{multirow}
\usepackage{amsmath}
\usepackage{amssymb}
\usepackage{graphicx}
\usepackage[pdfusetitle,
 bookmarks=true,bookmarksnumbered=false,bookmarksopen=true,bookmarksopenlevel=1,
 breaklinks=true,pdfborder={0 0 0},pdfborderstyle={},backref=false,colorlinks=true]
 {hyperref}
\hypersetup{
 citecolor=blue,urlcolor=blue,linkcolor=blue}

\makeatletter

\InputIfFileExists{t2aenc.def}{}{%
  \errmessage{File `t2aenc.def' not found: Cyrillic script not supported}}

\providecommand{\tabularnewline}{\\}

\usepackage{bm}
\allowdisplaybreaks

\makeatother

\begin{document}
\title{Production of $D^{(*)}\bar{D}^{(*)}$ near the thresholds in $e^{+}e^{-}$
annihilation}
\author{S.G. Salnikov}
\email{S.G.Salnikov@inp.nsk.su}

\author{A.I. Milstein}
\email{A.I.Milstein@inp.nsk.su}

\affiliation{\textit{Budker Institute of Nuclear Physics, 630090, Novosibirsk,
Russia}}
\affiliation{\textit{Novosibirsk State University, 630090, Novosibirsk, Russia}}
\date{\today}
\begin{abstract}
	It is shown that the nontrivial energy dependencies of $D\bar{D}$, $D\bar{D}^{*}$, and $D^{*}\bar{D}^{*}$ pair production cross sections in $e^{+}e^{- }$
	annihilation are well described within the  approach based on account for the final-state interaction of produced particles. This statement is
	valid for production of  charged and neutral particles. Interaction of $D^{(*)}$ and $\bar{D}^{(*)}$ is taken into account
	using the effective potential method. Its applicability is
	based on the fact that for near-threshold resonance the characteristic width of peak in the wave function is much larger than the interaction radius. The transition amplitudes between all three channels play an important role in the description of cross sections. These transitions are possible since all channels have the same quantum 	numbers $J^{PC}=1^{--}$.
\end{abstract}
\maketitle


\section{Introduction}
Currently, many dozens of resonances, having
very nontrivial energy dependence of the cross sections of processes, have been discovered:
 $e^{+}e^{-}\to p\bar{p}$~\citep{Aubert2006,Lees2013,Akhmetshin2016,Ablikim2020,Ablikim2021b,Akhmetshin2019,Ablikim2015,Ablikim2019},
$e^{+}e^{-}\to n\bar{n}$~\citep{Achasov2014,Ablikim2021f,Achasov2022},
$e^{+}e^{-}\to\Lambda\bar{\Lambda}$~\citep{Aubert2007,Ablikim2018,Ablikim2019c,Ablikim2023},
$e^{+}e^{-}\to\Lambda_{c}\bar{\Lambda}_{c}$~\citep{Pakhlova2008,Ablikim2018b,Ablikim2023Measurement},
$e^{+}e^{-}\to B\bar{B}$~\citep{Aubert2009,Mizuk2021}, and others.
In these processes, the widths of resonances  are of the order of
the distances to the thresholds of particle production, into which resonances mainly decay. In addition, the cross sections of production of light particles in the vicinity of
near-threshold resonances and the probabilities of heavy particle decays into certain channels also demonstrate a nontrivial energy dependence. For instance, such energy dependence is observed in the processes $e^{+}e^{-}\to6\pi$~\citep{Aubert2006a,Akhmetshin2013,Lukin2015,Akhmetshin2019},
$e^{+}e^{-}\to K^{+}K^{-}\pi^{+}\pi^{-}$~\citep{Aubert2005,Aubert2007c,Akhmetshin2019},
$J/\psi\to\gamma\eta'\pi^{+}\pi^{-}$~\citep{Ablikim2016}, and $J/\psi\to3\left(\pi^{+}\pi^{-}\right)\gamma$~\citep{Ablikim2023e}.
Despite the current availability of a fairly large amount of experimental data, the debate about the nature of near-threshold resonances is still ongoing.

Natural explanation of near-threshold resonances is based on account for the
interaction of produced particles. In this approach,
resonances arise in two cases (see, e.g.,~\citep{Salnikov2023,Salnikov2024}
and references therein). In the first case, there is a bound state with
the binding energy much less than the characteristic value of the interaction potential
(about several hundreds of MeV). In the second case, there is no loosely bound state
but a slight increase in the depth of potential leads to appearance of such state (this is the so-called virtual level).
In both cases, at scattering of produced hadrons  on each other, the modulus of scattering length   significantly exceeds the characteristic potential size (of the order of 1~fm). At the same time, the wave function at small distances calculated with  account for the interaction of produced  hadrons has characteristic value  much larger than that without account for the interaction. The ratio of squares of the modules of corresponding wave functions for the relative angular momentum $l=0$ (or their derivatives for $l=1$) is the
amplification factor, which can be very large. As a result,
resonant structures arise in the particle production cross section. Currently, more and more scientists are coming to the conclusion that taking into account the interaction in the final state is of crucial importance for the correct description of cross sections in the near-threshold region (see, e.g.,~\citep{Haidenbauer2024Bar} and references therein).

The description of final-state interaction becomes noticeably more complicated,
when there are several near-threshold resonances with the same quantum
numbers and thresholds located close to each other. As a result,
non-zero transition amplitudes between resonances arise, which leads to
to a significant distortion of the resonance shape. In our recent work~\citep{Salnikov2024}, we have  discussed various cases of coupled channels, where each channel is either loosely bound or virtual
state.  Moreover, it is shown in Ref.~\citep{Salnikov2024}  that the account for the final-state
interaction allows one to successfully describe the  $B^{(*)}\bar{B}^{(*)}$ production
near the thresholds in $e^{+}e^{-}$ annihilation. Similar results
for the system of $B^{(*)}\bar{B}^{(*)}$ mesons were obtained in Ref.~\citep{Husken2022Matrix}
using the $K$-matrix approach.

In this work, the processes $e^{+}e^{-}\to D^{(*)}\bar{D}^{(*)}$
near the thresholds are discussed.  Our approach is based on account for  the final-state interaction in the case of coupled channels. Certainly, our information on  the interaction potential
between $D$ mesons is very limited. However, it is not necessary to know these potentials very precisely.
As already mentioned, the characteristic size of a peak in the wave function of produced $D^{(*)}\bar{D}^{(*)}$
system near the threshold is much larger than the characteristic size of the potential. Therefore, specific shapes of the potentials are not important. They can be parameterized in any convenient way by few parameters. The numerical values of parameters are obtained by comparison of theoretical predictions and experimental data.

\section{ Description of the model.}
Pairs $D\bar{D}$, $D\bar{D}^{*}$, and $D^{*}\bar{D}^{*}$ are produced
in $e^{+}e^{-}$ annihilation in the states with quantum numbers $J^{PC}=1^{--}$.
In this case, the relative angular momentum of produced particles is
$l=1$. Due to $C$-parity conservation, the total spin $S$ of $D^{*}\bar{D}^{*}$ pair can
 be either $S=0$ or $S=2$. However, due to the lack of experimental
data for individual spin states in the $D^{*}\bar{D}^{*}$ channel,
we will talk on  the total cross section for the production of these states with different spins. At small distances $r\sim1/\sqrt{s}$, where $\sqrt{s}$
is the total energy of electron and positron in the center-of-mass frame,
a hadronic system is produced as $c\bar{c}$ pair and, therefore, has isospin $I=0$.
However, at large distances $r\gtrsim1/\Lambda_{QCD}$ the difference in
masses of charged and neutral $D$ mesons ($D^{*}$ mesons), as well as the Coulomb
interaction between charged particles,  leads to
violation of isospin  invariance. Thus, we have
six states with $C=-1$: $\Psi_{1}=D^{0}\bar{D}^{0}$, $\Psi_{2}=D^{+}D^{-}$,
$\Psi_{3}=(D^{0}\bar{D}^{0*}+\bar{D}^{0}D^{0*})/\sqrt{2}$, $\Psi_{4}=(D^{+}D^{-*}+D^{-}D^{+*})/\sqrt{2}$,
$\Psi_{5}=D^{0*}\bar{D}^{0*}$, and $\Psi_{6}=D^{+*}D^{-*}$.
Taking into account violation of isospin invariance, we conclude that it is necessary to solve the six-channel problem. The threshold of $\Psi_{1}$ state production is $3730\,\mbox{MeV}$.
We will count the remaining thresholds $\Delta_{i}$ from this value.
Therefore, $\Delta_{1}=0$, $\Delta_{2}=9.6\,\mbox{MeV}$, $\Delta_{3}=142\,\mbox{MeV}$,
$\Delta_{4}=150\,\mbox{MeV}$, $\Delta_{5}=284\,\mbox{MeV}$, and $\Delta_{6}=291\,\mbox{MeV}$.

The radial Schr\"odinger equation, which describes our six-channel system, has the form
\begin{equation}
\left(p_{r}^{2}+M_{D}\mathcal{V}+\frac{l(l+1)}{r^{2}}-\mathcal{K}^{2}\right)\bm{\Psi}(r)=0\,,\quad\left(\mathcal{K}^{2}\right)_{ij}=\delta_{ij}\,k_{i}^{2}\,,\quad\mathcal{V}=\begin{pmatrix}V_{11} & V_{12} & V_{13}\\
V_{12} & V_{22} & V_{23}\\
V_{13} & V_{23} & V_{33}
\end{pmatrix}\,,\label{eq}
\end{equation}
where $\left(-p_{r}^{2}\right)$ is the radial part of the Laplacian, $k_{i}=\sqrt{M_{D}\left(E-\Delta_{i}\right)}$,
$M_{D}=1865\,\mbox{MeV}$ is the  $D^{0}$ mass, $E$ is the energy of a system counted from the threshold of $D^0\bar{D}^0$ production, and $l=1$. The wave function
\[
\bm{\Psi}(r)=\left(\psi_{1}(r),\dots,\psi_{6}(r)\right)^{T}
\]
consists of radial parts $\psi_{i}(r)$ of wave functions of states
$\Psi_{i}$, index $T$
denotes transposition. The matrices $V_{ij}$ are symmetric
blocks of dimension $2\times2$ having the form
\begin{equation}
V_{ij}=\begin{pmatrix}U_{ij}^{(0)}(r)-U_{ij}^{(1)}(r) & -2U_{ij}^{(1)}(r)\\
-2U_{ij}^{(1)}(r) & U_{ij}^{(0)}(r)-U_{ij}^{(1)}(r)
\end{pmatrix}\,,\label{eq:pot1}
\end{equation}
where the diagonal potentials correspond to the transitions without change of
particle electric charges, and off-diagonal ones describe processes with charge exchange.
These potentials contain contributions from isoscalar and isovector exchange,
$U_{ij}^{(0)}(r)$ and $U_{ij}^{(1)}(r)$, respectively. All potentials
can be parameterized as
\begin{equation}
U_{ij}^{(I)}(r)=u_{ij}^{(I)}\,\theta(a_{ij}^{(I)}-r)\,.\label{eq:pot2}
\end{equation}
Here $\theta(x)$ is the Heaviside function, $u_{ij}^{(I)}$ and $a_{ij}^{(I)}$
are some constants that are found from comparison of theoretical predictions with experimental data.

The equation \eqref{eq} has six linearly independent regular at origin solutions,
\begin{equation}
\bm{\Psi}^{(m)}=\left(\psi_{1}^{(m)}(r),\dots,\psi_{6}^{(m)}(r)\right)^{T},\qquad m=1,\dots,6\,.
\end{equation}
Each solution is determined by the asymptotic behavior at $r\rightarrow\infty$,
\begin{align}
 & \bm{\Psi}^{(m)}=\frac{1}{2ik_{m}r}\left(S_{1}^{(m)}\chi_{1}^{+},\dots,S_{m}^{(m)}\chi_{m}^{+}-\chi_{m}^{-},\dots,S_{6}^{(m)}\chi_{6}^{+}\right)^{T},\nonumber \\
 & \chi_{i}^{\pm}=\exp\left[\pm i\left(k_{i}r-\pi/2\right)\right],
\end{align}
where $S_{i}^{(m)}$ are some coefficients. The cross sections $\sigma^{(m)}$ of pair production in the states~$\Psi_{m}$ have the form
\begin{align}
 & \sigma^{(m)}=\frac{2\pi\beta_{m}\alpha^{2}}{s}\left|\sum_{i=1}^{6}g_{i}\dot{\psi}_{i}^{(m)}(0)\right|^{2}\,.\label{sec}
\end{align}
Here $\beta_{m}=k_{m}/M_{D}$, $g_{i}$ are some constants that determine
the production  of corresponding states at small distances, $\dot{\psi}_{i}^{(m)}(r)=\partial/\partial r\,\psi_{i}^{(m)}(r)$.
Since an isoscalar state is produced at small distances, then
$g_{1}=g_{2}$, $g_{3}=g_{4}$, and $g_{5}=g_{6}$.

\section{Results}

In Refs.~\citep{Aubert2007Study,Pakhlova2007Measurement,Pakhlova2008Measurement,Cronin-Hennessy2009Measurement,Dong2018Derived,Zhukova2018Angular,Ablikim2022Cross,Ablikim2024Precise}
detailed experimental data on cross sections $\sigma^{(m)}$ have been obtained
for all~$m$. Parameters $u_{ij}^{(I)}$, $a_{ij}^{(I)}$,
and $g_{i}$ of our model are determined by comparing predictions with
all experimental data listed above. We analyze
data for energies $E$ up to $\unit[450]{MeV}$ since, on the one hand, we want to cover the range of thresholds of all six channels, and on the other hand, we use a non-relativistic model and cannot consider too high energies.

\setlength{\tabcolsep}{1em}
\renewcommand{\arraystretch}{1.5}
\begin{table}
\begin{center}
\begin{tabular}{|l|c|c|c|c|}
\hline
\multirow{2}{*}{} & \multicolumn{2}{c|}{Isoscalar exchange} & \multicolumn{2}{c|}{Isovector exchange}\tabularnewline
\cline{2-5}
 & $u^{(0)}\,(\textrm{MeV})$ & $a^{(0)}\,(\textrm{fm})$ & $u^{(1)}\,(\textrm{MeV})$ & $a^{(1)}\,(\textrm{fm})$\tabularnewline
\hline
$V_{11}$ & $-233.2$ & $1.432$ & $56.5$ & $1.925$\tabularnewline
$V_{22}$ & $-104$ & $1.61$ & $184.6$ & $0.932$\tabularnewline
$V_{33}$ & $-18.4$ & $2.198$ & $129.5$ & $1.263$\tabularnewline
$V_{12}$ & $143.5$ & $1.708$ & $5.9$ & $2.671$\tabularnewline
$V_{13}$ & $43$ & $1.73$ & $-100.9$ & $0.443$\tabularnewline
$V_{23}$ & $-22.5$ & $1.821$ & $-13.6$ & $1.425$\tabularnewline
\hline
\end{tabular}
\par\end{center}
\caption{Parameters of interaction potentials defined by Eqs.~\eqref{eq:pot1}
and~\eqref{eq:pot2}.}\label{tab:pot}
\end{table}

\begin{figure}
\begin{center}
\includegraphics[totalheight=6cm]{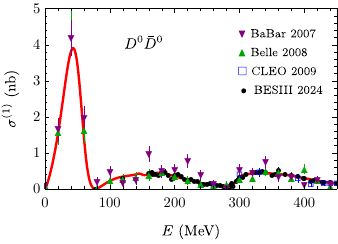}\hfill{}\includegraphics[totalheight=6cm]{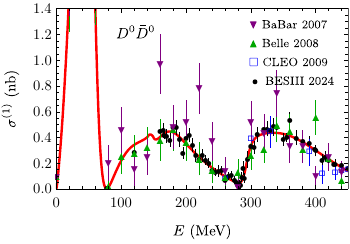}
\par\end{center}
\centering{}\includegraphics[totalheight=6cm]{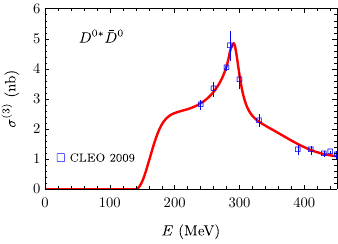}\hfill{}\includegraphics[totalheight=6cm]{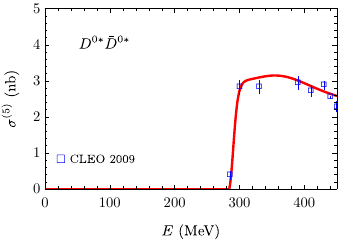}
\caption{Energy dependence of the cross sections for the production of neutral  particles. Experimental
	data are taken from Refs.~\citep{Aubert2007Study,Pakhlova2008Measurement,Cronin-Hennessy2009Measurement,Dong2018Derived,Ablikim2024Precise}.}\label{fig:00}
\end{figure}

\begin{figure}
\begin{center}
\includegraphics[totalheight=6cm]{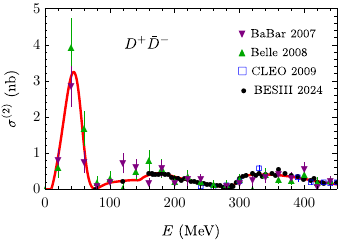}\hfill{}\includegraphics[totalheight=6cm]{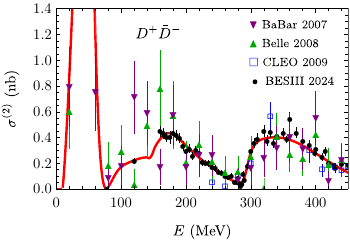}
\par\end{center}
\centering{}\includegraphics[totalheight=6cm]{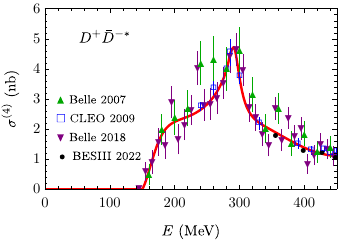}\hfill{}\includegraphics[totalheight=6cm]{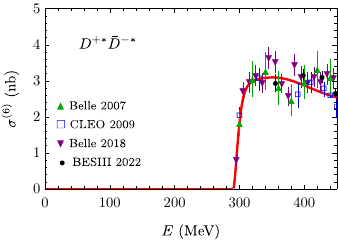}\caption{
	Energy dependence of the cross sections for production of charged particles. Experimental
	data are taken from Refs.~\citep{Aubert2007Study,Pakhlova2007Measurement,Pakhlova2008Measurement,Cronin-Hennessy2009Measurement,Dong2018Derived,Zhukova2018Angular,Ablikim2022Cross,Ablikim2024Precise}.}\label{fig:pm}
\end{figure}

The  parameters of the model are obtained using the $\chi^{2}$ minimization method.
The values of parameters  that provide the best agreement with experiment
are given in Table~\ref{tab:pot}. Constants, that determine production of
different states at small distances, have the values $g_{1}=g_{2}=0.069$,
$g_{3}=g_{4}=0.003+0.169\,i$, and $g_{5}=g_{6}=0.429-0.156\,i$. As a result of
fitting, we have obtained $\chi^{2}/N_{\textrm{df}}=325/301=1.08$, where
$N_{\textrm{df}}$ is the number of degrees of freedom.  The latter  equals  to the difference
between the number of experimental points and the number of parameters in the model.

Figs.~\ref{fig:00} and~\ref{fig:pm} show a comparison of our theoretical
predictions with experimental data from Refs.~\citep{Aubert2007Study,Pakhlova2007Measurement,Pakhlova2008Measurement,Cronin-Hennessy2009Measurement,Dong2018Derived,Zhukova2018Angular,Ablikim2022Cross,Ablikim2024Precise}.
It is seen that good agreement of predictions with experimental data is obtained over the entire energy range under consideration. In particular, recent data from Ref.~\citep{Ablikim2024Precise} is perfectly described by our model.
Few experimental points lie outside of our theoretical predictions, but this is related to the fact that these points have large experimental uncertainties and are not consistent with each other.

The cross sections  of different $D$ meson pair  production have very
non-trivial energy dependencies. There are  many peaks
of various shapes, as well as sharp gaps between them. Note that  experimental data obtained for all six charged and neutral channels have high accuracy. Therefore, for simultaneous description of the
cross sections of these processes, it is necessary to take into account all six  channels
and all possible transitions between them.
All potentials (diagonal and off-diagonal, with charge exchange and without charge exchange)
are important to obtained good agreement between theory and experiment.

\section{Conclusion}

It is shown that the final-state interaction in the system of
$D^{(*)}$ mesons  explains the nontrivial energy dependence of the
cross sections of  $e^{+}e^{-}\to D^{(*)}\bar{D}^{(*)}$ annihilation. Interaction between
$D^{(*)}$ mesons is described using the effective potentials. Their parameters
 are determined from comparison of experimental data with theoretical predictions in each channel. Good agreement is  obtained for the cross sections of  charged and neutral pair production. We emphasize again that, to obtain a good description of experimental data for the processes $e^{+}e^{-}\to D^{(*)}\bar{D}^{(*)}$, it is necessary to take into account all six channels simultaneously and all transitions between them.

Quite recently, a work~\citep{Husken2024Poles} has appeared, where the cross sections of
processes $e^{+}e^{-}\to D^{(*)}\bar{D}^{(*)}$ have been described using
$K$-matrix approach. Although the approach of Ref.~\citep{Husken2024Poles}
differs significantly from ours and the experimental data averaged over isospin are used in the channels $D^{*}\bar{D}$ and $D^{*}\bar{D}^{*}$, the results of Ref.~\citep{Husken2024Poles} are consistent with ours qualitatively and quantitatively.

\end{document}